\PassOptionsToPackage{unicode}{hyperref}
\PassOptionsToPackage{hyphens}{url}
\PassOptionsToPackage{dvipsnames,svgnames,x11names}{xcolor}
\documentclass[
]{article}

\usepackage{amsmath,amssymb}
\usepackage{iftex}
\ifPDFTeX
  \usepackage[T1]{fontenc}
  \usepackage[utf8]{inputenc}
  \usepackage{textcomp} % provide euro and other symbols
\else % if luatex or xetex
  \usepackage{unicode-math}
  \defaultfontfeatures{Scale=MatchLowercase}
  \defaultfontfeatures[\rmfamily]{Ligatures=TeX,Scale=1}
\fi
\usepackage{lmodern}
\ifPDFTeX\else  
\fi
\IfFileExists{upquote.sty}{\usepackage{upquote}}{}
\IfFileExists{microtype.sty}{% use microtype if available
  \usepackage[]{microtype}
  \UseMicrotypeSet[protrusion]{basicmath} % disable protrusion for tt fonts
}{}
\makeatletter
\@ifundefined{KOMAClassName}{% if non-KOMA class
  \IfFileExists{parskip.sty}{%
    \usepackage{parskip}
  }{% else
    \setlength{\parindent}{0pt}
    \setlength{\parskip}{6pt plus 2pt minus 1pt}}
}{% if KOMA class
  \KOMAoptions{parskip=half}}
\makeatother
\usepackage{xcolor}
\makeatletter
\ifx\paragraph\undefined\else
  \let\oldparagraph\paragraph
  \renewcommand{\paragraph}{
    \@ifstar
      \xxxParagraphStar
      \xxxParagraphNoStar
  }
  \newcommand{\xxxParagraphStar}[1]{\oldparagraph*{#1}\mbox{}}
  \newcommand{\xxxParagraphNoStar}[1]{\oldparagraph{#1}\mbox{}}
\fi
\ifx\subparagraph\undefined\else
  \let\oldsubparagraph\subparagraph
  \renewcommand{\subparagraph}{
    \@ifstar
      \xxxSubParagraphStar
      \xxxSubParagraphNoStar
  }
  \newcommand{\xxxSubParagraphStar}[1]{\oldsubparagraph*{#1}\mbox{}}
  \newcommand{\xxxSubParagraphNoStar}[1]{\oldsubparagraph{#1}\mbox{}}
\fi
\makeatother

\usepackage{longtable,booktabs,array}
\usepackage{calc} % for calculating minipage widths
\usepackage{etoolbox}
\makeatletter
\patchcmd\longtable{\par}{\if@noskipsec\mbox{}\fi\par}{}{}
\makeatother
\IfFileExists{footnotehyper.sty}{\usepackage{footnotehyper}}{\usepackage{footnote}}
\makesavenoteenv{longtable}
\usepackage{graphicx}
\makeatletter
\def\maxwidth{\ifdim\Gin@nat@width>\linewidth\linewidth\else\Gin@nat@width\fi}
\def\maxheight{\ifdim\Gin@nat@height>\textheight\textheight\else\Gin@nat@height\fi}
\makeatother
\setkeys{Gin}{width=\maxwidth,height=\maxheight,keepaspectratio}
\makeatletter
\def\fps@figure{htbp}
\makeatother
\NewDocumentCommand\citeproctext{}{}

\makeatletter
 \let\@cite@ofmt\@firstofone
 \def\@biblabel#1{}
 \def\@cite#1#2{{#1\if@tempswa , #2\fi}}
\makeatother
\newlength{\cslhangindent}
\newlength{\csllabelwidth}
\newenvironment{CSLReferences}[2] % #1 hanging-indent, #2 entry-spacing
 {\begin{list}{}{%
  \setlength{\itemindent}{0pt}
  \setlength{\leftmargin}{0pt}
  \setlength{\parsep}{0pt}
  \ifodd #1
   \setlength{\leftmargin}{\cslhangindent}
   \setlength{\itemindent}{-1\cslhangindent}
  \fi
  \setlength{\itemsep}{#2\baselineskip}}}
 {\end{list}}
\usepackage{calc}

\usepackage{arxiv}
\usepackage{orcidlink}
\usepackage{amsmath}
\usepackage[T1]{fontenc}
\makeatletter
\@ifpackageloaded{caption}{}{\usepackage{caption}}
\AtBeginDocument{%
\ifdefined\contentsname
  \renewcommand*\contentsname{Table of contents}
\else
  \newcommand\contentsname{Table of contents}
\fi
\ifdefined\listfigurename
  \renewcommand*\listfigurename{List of Figures}
\else
  \newcommand\listfigurename{List of Figures}
\fi
\ifdefined\listtablename
  \renewcommand*\listtablename{List of Tables}
\else
  \newcommand\listtablename{List of Tables}
\fi
\ifdefined\figurename
  \renewcommand*\figurename{Figure}
\else
  \newcommand\figurename{Figure}
\fi
\ifdefined\tablename
  \renewcommand*\tablename{Table}
\else
  \newcommand\tablename{Table}
\fi
}
\@ifpackageloaded{float}{}{\usepackage{float}}
\floatstyle{ruled}
\@ifundefined{c@chapter}{\newfloat{codelisting}{h}{lop}}{\newfloat{codelisting}{h}{lop}[chapter]}
\floatname{codelisting}{Listing}

\makeatother
\makeatletter
\@ifpackageloaded{caption}{}{\usepackage{caption}}
\@ifpackageloaded{subcaption}{}{\usepackage{subcaption}}
\makeatother

\ifLuaTeX
  \usepackage{selnolig}  % disable illegal ligatures
\fi
\usepackage{bookmark}

\IfFileExists{xurl.sty}{\usepackage{xurl}}{} % add URL line breaks if available
\hypersetup{
  pdftitle={Sensitivity-aware rock physics enhanced digital shadow for underground-energy storage monitoring},
  pdfauthor={Abhinav Prakash Gahlot; Huseyin Tuna Erdinc; Felix J. Herrmann},
  colorlinks=true,
  linkcolor={blue},
  filecolor={Maroon},
  citecolor={Blue},
  urlcolor={Blue},
  pdfcreator={LaTeX via pandoc}}

\newcommand{\runninghead}{A Preprint }
\title{Sensitivity-aware rock physics enhanced digital shadow for
underground-energy storage monitoring}
\def\asep{\\\\\\ } % default: all authors on same column
\author{\textbf{Abhinav Prakash Gahlot}\\\\Georgia Institute of
Technology\\\\\asep\textbf{Huseyin Tuna Erdinc}\\\\Georgia Institute of
Technology\\\\\asep\textbf{Felix J. Herrmann}\\\\Georgia Institute of
Technology\\\\}
\date{}
\begin{document}
\maketitle
\begin{abstract}
Underground energy storage, which includes storage of hydrogen,
compressed air, and CO\textsubscript{2}, requires careful monitoring to
track potential leakage pathways, a situation where time-lapse seismic
imaging alone may be inadequate. A recently developed Digital Shadow
(DS) enhances forecasting using machine learning and Bayesian inference,
yet their accuracy depends on assumed rock physics models, the mismatch
of which can lead to unreliable predictions for the reservoir's state
(saturation/pressure). Augmenting DS training with multiple rock physics
models mitigates errors but averages over uncertainties, obscuring their
sources. To address this challenge, we introduce context-aware
sensitivity analysis inspired by amortized Bayesian inference, allowing
the DS to learn explicit dependencies between seismic data, the
reservoir state, e.g., CO\textsubscript{2} saturation, and rock physics
models. At inference time, this approach allows for real-time ``what
if'' scenario testing rather than relying on costly retraining, thereby
enhancing interpretability and decision-making for safer, more reliable
underground storage.
\end{abstract}

\newcommand{\argmin}{\mathop{\mathrm{argmin}\,}\limits}
\newcommand{\argmax}{\mathop{\mathrm{argmax}\,}\limits}

\[
\def\textsc#1{\dosc#1\csod} 
\def\dosc#1#2\csod{{\rm #1{\small #2}}} 
\]

\section{INTRODUCTION}\label{introduction}

The transition to a more sustainable and resilient energy system
requires efficient underground storage solutions for energy carriers
such as hydrogen, compressed air, and CO\textsubscript{2}. As global
energy demand continues to rise, the intermittency of renewable sources
and the need for a stable energy supply necessitate large-scale storage
technologies (Global CCS Institute (GCCSI) 2019; Ringrose 2020).
Subsurface storage plays a crucial role in balancing energy supply and
demand by providing buffer capacity for excess renewable energy and
supporting decarbonization efforts in power generation, manufacturing,
and transportation (International Energy Agency (IEA) 2016). Geological
storage enables the safe and long-term containment of these energy
carriers while ensuring operational reliability and energy security
(Ringrose 2020; IPCC special report 2018). However, large-scale
deployment requires advancements in storage integrity monitoring,
subsurface flow characterization, and leakage risk mitigation (Ringrose
2023). Precise monitoring of subsurface dynamics, particularly in
complex multi-phase flow environments, is essential to maintaining
storage efficiency and safety. While geophysical techniques such as
time-lapse seismic imaging offer valuable insights, they often lack the
resolution needed to fully capture the intricacies of underground energy
storage processes.

Digital Shadows (DS), powered by machine learning-driven data
assimilation techniques such as nonlinear Bayesian filtering and
generative AI (Spantini, Baptista, and Marzouk 2022; Gahlot et al.
2024), offer a high-fidelity approach to characterize subsurface
CO\textsubscript{2} flow (Herrmann 2023; Gahlot et al. 2023, 2024). By
incorporating uncertainty in reservoir properties such as permeability,
DS provides uncertainty-aware CO\textsubscript{2} migration forecasts,
including predictions of plume pressure and saturation, thereby reducing
risks in geological storage (GCS) projects by enabling decision-making
under uncertainty.

\begin{figure}

\begin{minipage}{\linewidth}

\includegraphics[width=1\textwidth,height=\textheight]{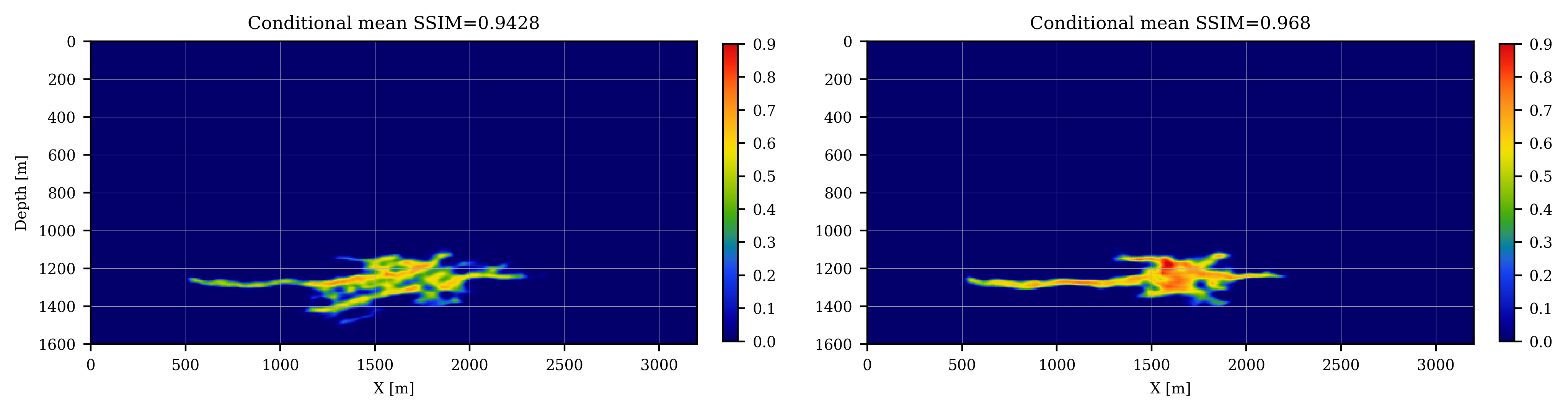}

\end{minipage}%

\caption{\label{fig-DS-breaks}Conditional mean obtained when DS is
conditioned on data obtained from uniform (left) and patchy model
(right)}

\end{figure}

However, the accuracy of these forecasts depends on assumptions
regarding reservoir properties, rock physics models, and initial
conditions. If these assumptions are inaccurate, predictions can become
unreliable, compromising storage safety. figure~\ref{fig-DS-breaks}
illustrates what can happen when a Digital Shadows (DS) model is trained
using a patchy rock physics model and evaluated on data generated with a
uniform rock physics model, in which case it produces an incorrect plume
prediction (left), as compared to the ground-truth plume shown in the
right plot of figure~\ref{fig-GT}. However, when the DS is conditioned
on time-lapse seismic data generated with the patchy model, it yields
the correct plume prediction. The similarity between these predicted
plumes and the ground truth is quantified using the Structural
Similarity Index (SSIM), which is higher in the right plot than on the
left, indicating better agreement with the ground truth when the correct
rock physics model is used. To address this challenge, Gahlot and
Herrmann (2025) proposed augmenting the forecast ensemble used for
training neural networks in the data assimilation process. By
incorporating multiple rock physics models ranging from patchy to
uniform saturation, their approach mitigates the impact of model
misspecification and enhances predictive accuracy in key scenarios.
While this approach marginalizes over plausible rock physics models, it
sacrifices the ability to reason explicitly about the influence of
individual rock physics. By producing a marginal posterior, the plume
predictions become averaged across all possible rock physics models,
resulting in broader uncertainty bounds, which can obscure the true
source of uncertainty and complicate decision-making. Without an
explicit treatment of rock physics variability, it becomes unclear
whether uncertainties arise due to seismic noise, uncertain
permeability, or incorrect assumptions about the rock physics model
itself. This lack of interpretability can reduce confidence in
monitoring and forecasting CO\textsubscript{2} plume behavior in
geological storage settings.

To circumvent this, we need a framework that enables sensitivity
analysis of the Digital Shadow over a range of rock physics models.
However, training a separate network for each rock physics model is
computationally infeasible, as retraining is costly and data-intensive.
Traditionally, sensitivity analyses have relied on computationally
expensive refitting procedures, where a new model is trained for each
specific configuration, limiting scalability and generalization towards
real-world monitoring scenarios. Elsemüller et al. (2024) demonstrates
that Amortized Bayesian Inference (ABI) (Radev et al. 2020) provides a
principled way to assess the sensitivity of model predictions without
requiring separate network training for each context. ABI leverages
weight sharing within a single network to generalize across different
contexts, allowing for computationally efficient inference on unseen
datasets. Additionally, incorporating context variables into the neural
network, ABI facilitates systematic uncertainty quantification and
enables the model to learn a context-aware conditional posterior that
captures how inferred parameters depend on varying forward model
assumptions. This allows for a faster and more structured approach to
sensitivity analysis, where the impact of different rock physics models,
in the case of DS, can be assessed directly at inference time rather
than through costly retraining.

Motivated by this, we extend the approach in Gahlot and Herrmann (2025)
by incorporating explicit rock physics sensitivity analysis within the
DS framework. Instead of training separate models for different rock
physics configurations, we introduce context variables representing the
rock physics model into the training process. This allows the network to
learn an explicit mapping between seismic data, CO\textsubscript{2}
saturation, and the rock physics model, thereby improving robustness,
interpretability, and generalizability. By including rock physics as a
context variable, we perturb seismic images corresponding to
CO\textsubscript{2} saturation forecasts in the training ensembles in a
structured manner. This ensures that the DS can dynamically adapt to
different rock physics models during inference rather than producing a
posterior smeared over all possible models. As a result, our approach
enables not only uncertainty quantification over CO\textsubscript{2}
plume forecasts but also a direct assessment of how different
assumptions on the rock physics model affect CO\textsubscript{2} plume
prediction results, enhancing decision-making for underground-energy
storage monitoring.

\section{METHODOLOGY}\label{methodology}

Building upon the uncertainty-aware Digital Shadow (DS) framework
introduced in (Gahlot et al. 2024), we develop a Bayesian
inference-driven approach for tracking the storage reservoir's state.
The temporal dynamics of this state, e.g., the CO\textsubscript{2}
saturation and pressure perturbations, are modeled as:

\begin{equation}\phantomsection\label{eq-dynamics}{
\begin{aligned}
\mathbf{x}_k & = \mathcal{M}_k\bigl(\mathbf{x}_{k-1}, \boldsymbol{\kappa}_k\bigr), \ \boldsymbol{\kappa}_k \sim p(\boldsymbol{\kappa}) \quad \text{for}\quad k=1, \dots, K.
\end{aligned}
}\end{equation}

where \(\mathbf{x}_k\) represents the reservoir's state at time step
\(k\), governed by the multi-phase fluid-flow operator
\(\mathcal{M}_k\). Because of lack of knowledge on the static reservoir
properties, the permeability field \(\boldsymbol{\kappa}\), is
inherently uncertain due to subsurface heterogeneity (Ringrose 2020). To
account for this uncertainty, we model the permeability as a random
field drawn from the probability distribution,
\(p(\boldsymbol{\kappa})\), each time the fluid-flow operator is
applied. This ensures that the simulations generate a diverse set of
plausible scenarios, capturing the inherent variability of the
reservoir.

While fluid-flow simulations provide physically consistent dynamics,
they remain impractical without observational constraints due to the
inherent stochasticity of the permeability and the uncertainty in the
employed rock physics models tying the reservoir's state to time-lapse
seismic data. Time-lapse seismic imaging (Lumley 2010) serves as a key
monitoring tool, providing indirect observations of the reservoir's
state. The corresponding observation model is defined as:

\begin{equation}\phantomsection\label{eq-obs}{
\mathbf{y}_k = \mathcal{H}_k(\mathbf{x}_k;\mathcal{R}_k,\boldsymbol{\epsilon}_k), \quad \boldsymbol{\epsilon}_k \sim p(\boldsymbol{\epsilon}), \quad \mathcal{R}_k \sim p(\mathcal{R}) \quad \text{for}\quad k=1, \dots, K
}\end{equation}

where \(\mathbf{y}_k\) represents seismic images derived from shot data
recorded at timestep, \(t_k\), \(\mathcal{H}_k\) is the seismic
observation operator, and \(\boldsymbol{\epsilon}_k\) is the colored
Gaussian noise added to the seismic shot records before reverse-time
migration, accounting for measurement errors. The term \(\mathcal{R}_k\)
corresponds to the rock physics, which links the fluid-flow properties
to the seismic properties. To account for uncertainty in rock physics,
we sample \(\mathcal{R}_k\) from a family of Brie Saturation models
(Avseth, Mukerji, and Mavko 2010), with the exponent \(e\) drawn from
the uniform distribution, \(e \sim \mathcal{U}(1,10)\). Including this
exponent entails a context-aware data augmentation strategy that ensures
that seismic images reflect a spectrum of plausible rock physics
assumptions. To infer the posterior distribution of the reservoir's
state from seismic data and the context, we employ Conditional
Normalizing Flows (CNFs, (Papamakarios et al. 2021; Gahlot et al. 2024))
within a simulation-based inference framework utilizing the training
pairs of reservoir states and corresponding seismic images by varying
both permeability and rock physics models (cf.~equations
\ref{eq-dynamics} and \ref{eq-obs}). This results in an enriched
augmented training set, where multiple seismic realizations are
generated along with their contexts for each sample of the state. The
CNF approximates the posterior distribution
\(p(\mathbf{x}_k \mid \mathbf{y}_k)\), allowing for Bayesian updates as
new seismic data become available. This approach couples sequential
Bayesian inference with neural posterior density estimation, leveraging
deep generative models for rock physics context-aware forecasting of the
reservoir's state.

\section{SYNTHETIC CASE STUDY}\label{synthetic-case-study}

To assess the proposed methodology, we utilize a synthetic 2D Earth
model derived from the Compass model (E. Jones et al. 2012), which is
representative of geological formations in the North Sea region. A
subset of this model, containing key subsurface structures suitable for
CO\textsubscript{2} injection, is selected and discretized into a
computational grid of \(512 \times 256\) with a spatial resolution of
\(6.25 \ \mathrm{m}\). Initialization of the DS requires an ensemble of
potential CO\textsubscript{2} plume scenarios, which depend on the
inherent uncertainty in the permeability distribution of the storage
reservoir. To account for this variability, we establish a probabilistic
baseline velocity model, inferred through full-waveform inference (Yin
et al. 2024) under the assumption of a baseline seismic survey conducted
before CO\textsubscript{2} injection. The resulting samples of the
velocity distribution are then converted into permeability samples using
an empirical transformation elucidated in Gahlot et al. (2024).

\subsection{Multi-phase flow
simulations}\label{multi-phase-flow-simulations}

Flow simulations are conducted using the open-source tool JutulDarcy
JutulDarcy.jl (Møyner, Bruer, and Yin 2023). In the initial setup, the
reservoir is filled with brine, and supercritical CO\textsubscript{2} is
injected at a constant rate of \(0.0500 \ \mathrm{m^3/s}\) for 1920
days. The injection occurs at an approximate depth of
\(1200 \ \mathrm{m}\). The simulation is performed over four time-lapse
intervals, denoted as \(t_k\), generating predicted CO\textsubscript{2}
saturations for each of the \(N=128\) ensemble members at every
timestep.

\subsection{Context-augmented seismic
simulations}\label{context-augmented-seismic-simulations}

The outputs from the 128 flow simulations are translated into changes in
subsurface acoustic properties through the application of 6 different
contexts (exponent of the Brie Saturation model). This approach
effectively augments the dataset by a factor of six, generating distinct
acoustic changes for each seismic simulation. The seismic surveys are
conducted using 8 receivers and 200 sources, with a dominant frequency
of 15 Hz and a recording duration of 1.8 seconds. To simulate real-world
conditions, 28 dB SNR colored Gaussian noise is added to the shot
records. Nonlinear wave simulations and imaging are performed using the
open-source package
\href{https://github.com/slimgroup/JUDI.jl}{JUDI.jl}(Witte et al. 2019;
Louboutin et al. 2023), after incorporating the various rock physics
models.

\subsection{CNF training}\label{cnf-training}

The training dataset consists of \(768\) ensemble members, each
consisting of CO\textsubscript{2} plume forecasts, its corresponding
simulated seismic observations, and the context. A Conditional
Normalizing Flow (CNF) is trained using the open-source package
InvertibleNetworks.jl (Orozco et al. 2024). Inspired by the literature
on generative models (Song et al. 2020), we integrate the context not
through an additional channel but directly into each activation layer of
the conditional arm of CNF. Specifically, the randomly drawn Brie
exponent, \(e\), is first passed through a sinusoidal function (Tancik
et al. 2020), followed by a dense layer, and the resulting embedded
output is added to the intermediate activations of the conditional arm.
Our experiments across different implementation settings indicate that
this architectural design enhances the model's sensitivity to contextual
variations without compromising predictive accuracy. The context,
\(C = \mathbf{g}_{\boldsymbol{\gamma}}(e)\), corresponds to the
embedding of the Brie saturation exponent, \(e\), with learnable
function \(\mathbf{g}_{\boldsymbol{\gamma}}\) with network parameters
\(\boldsymbol{\gamma}\). We integrate this context into the standard
negative log-posterior objective of the CNF, represented as,
\(\mathbb{E}_{\mathbf{x} \sim q_{\phi}(\mathbf{x} | \mathbf{y}, \mathbf{g}_{\boldsymbol{\gamma}}(e))} \left[ - \log q_{\phi}(\mathbf{x} | \mathbf{y}, \mathbf{g}_{\boldsymbol{\gamma}}(e)) \right]\),
and is evaluated using the samples of the approximate surrogate
distribution \(q_{\phi}\). To achieve the required amortization over a
set of context variables \(C\), which is obtained from the Brie exponent
\(e \sim \mathcal{U}(1,10)\), we minimize the context-aware (CA) loss,
\(\mathbb{E}_{e \sim p(e)} \left[ \mathbb{E}_{(\mathbf{x} \sim q_{\phi}(\mathbf{x} | \mathbf{y}, \mathbf{g}_{\boldsymbol{\gamma}}(e))} \left[ - \log q_{\phi}(\mathbf{x} | \mathbf{y}, \mathbf{g}_{\boldsymbol{\gamma}}(e)) \right] \right]\),
where the outer expectation is evaluated using the samples of \(e\).
Following Gahlot et al. (2024), this corresponds to training a CNF where
the network parameters \(\boldsymbol{\phi}\) are optimized by minimizing
the following objective function over 300 epochs, utilizing the
\(\textsc{ADAM}\) optimizer (Kingma and Ba 2014):

\begin{equation}\phantomsection\label{eq-loss-CNF}{
\widehat{\boldsymbol{\phi}} = \mathop{\mathrm{argmin}\,}\limits_{\boldsymbol{\phi}} \frac{1}{M}\sum_{m=1}^M \Biggl(\frac{\Big\|f_{\boldsymbol{\phi}}(\mathbf{x}^{(m)};(\mathbf{y}^{(m)},C^{(m)}))\Big\|_2^2}{2} - \log\Bigl |\det\Bigl(\mathbf{J}^{(m)}_{f_{\boldsymbol{\phi}}}\Bigr)\Bigr |\Biggr).
}\end{equation}

where \(\mathbf{J}\) is the Jacobian of the network \(f_{\theta}\) with
respect to its input, and \(M\) is the number of training samples. For
further details, we refer to (Gahlot et al. 2024).

\section{RESULTS}\label{results}

\begin{figure}

\begin{minipage}{\linewidth}

\includegraphics[width=1\textwidth,height=\textheight]{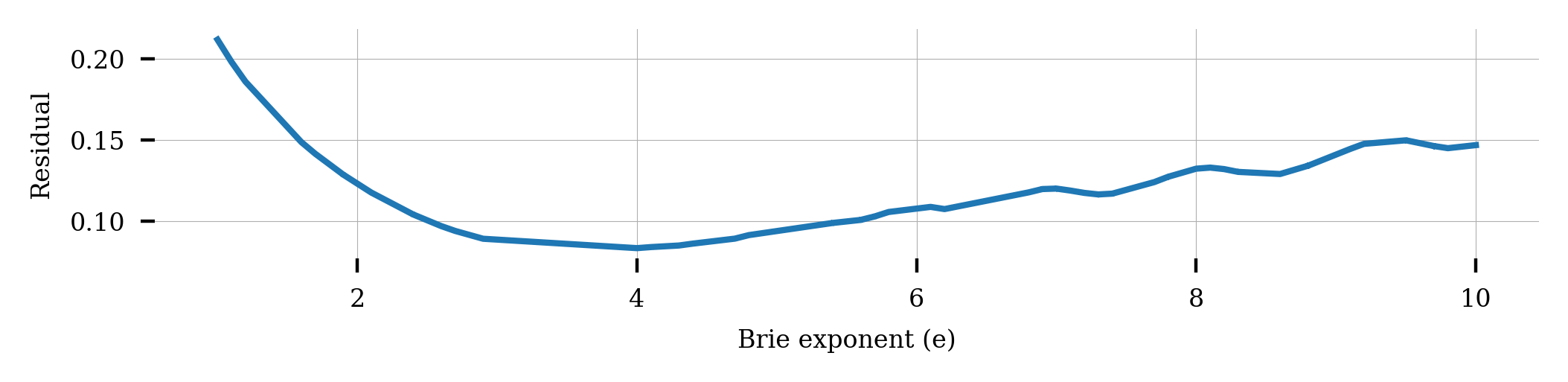}

\end{minipage}%

\caption{\label{fig-metrics}Data residual plot for various Brie
exponents}

\end{figure}

\begin{figure}

\begin{minipage}{\linewidth}

\includegraphics[width=1\textwidth,height=\textheight]{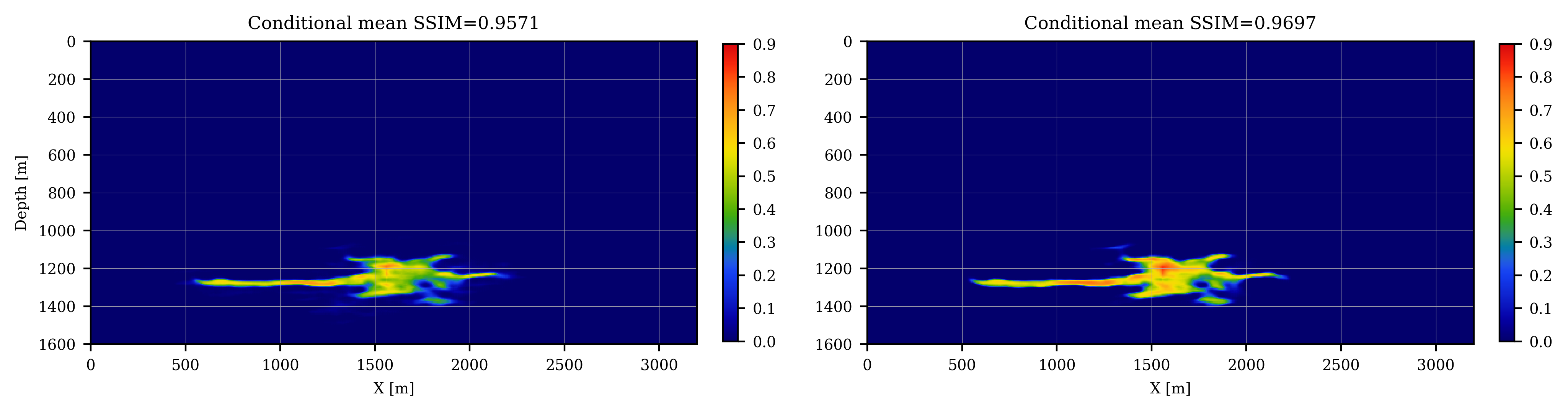}

\end{minipage}%

\caption{\label{fig-cond-mean}Conditional mean obtained from the trained
DS for incorrect (left) and correct Brie exponent (right)}

\end{figure}

\begin{figure}

\begin{minipage}{\linewidth}

\includegraphics[width=1\textwidth,height=\textheight]{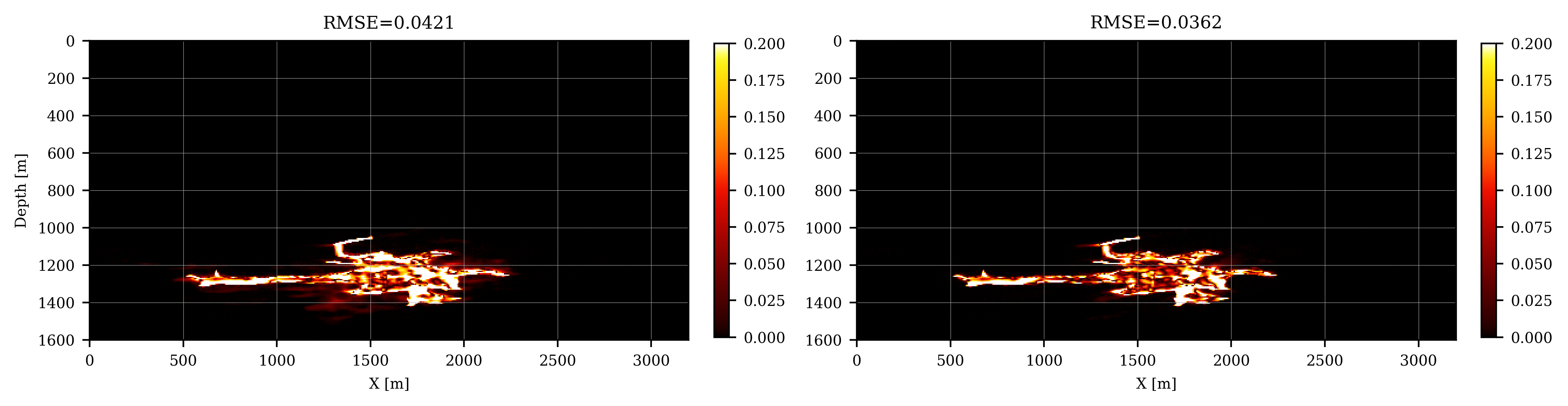}

\end{minipage}%

\caption{\label{fig-rmse}Error between conditional mean and ground truth
for incorrect (left) and correct Brie exponent (right)}

\end{figure}

\begin{figure}

\begin{minipage}{\linewidth}

\includegraphics[width=1\textwidth,height=\textheight]{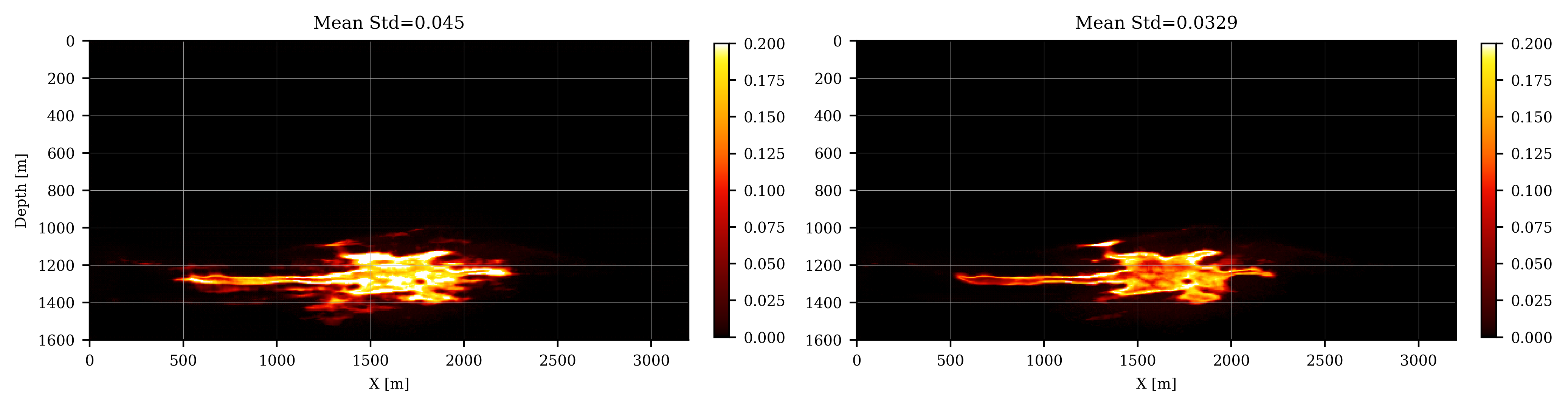}

\end{minipage}%

\caption{\label{fig-std}Standard deviation of DS posterior samples for
incorrect (left) and correct Brie exponent (right)}

\end{figure}

The performance of the context-aware DS framework, which incorporates
sensitivity to the chosen rock physics model, is presented below.
Figure~\ref{fig-GT} shows the observed time-lapse data (left) alongside
the ground-truth CO\textsubscript{2} plume (right).
Figure~\ref{fig-metrics} depicts the residual data, calculated as the
\(\ell_2\)-norm of the difference between the simulated data (obtained
by sampling various Brie exponents) and the ground-truth data. As
expected, the \(\ell_2\)-norm of the residual reaches its minimum at
\(e=4\), as this is the exponent used to generate the observed data.
Figure~\ref{fig-cond-mean} presents the conditional mean, where DS is
conditioned on time-lapse data generated with an incorrect exponent
(\(e=9\), left), which results in a low Structural Similarity Index
(SSIM). In contrast, conditioning on the correct exponent (\(e=4\),
right) yields a significantly higher SSIM. The captions of
Figure~\ref{fig-rmse} and figure~\ref{fig-std} show the Root Mean Square
Error (RMSE) and mean standard deviation (std) values, respectively. The
incorrect exponent (\(e=9\), left) leads to higher RMSE and mean std
values, while the correct exponent (\(e=4\), right) results in lower
values for both. Additionally, the left plot in figure~\ref{fig-rmse}
highlights a larger error at the bottom of the plume due to conditioning
on the incorrect exponent, whereas this error is notably reduced in the
right plot when conditioning on the correct exponent. Similarly, the
standard deviation is considerably higher in the left plot and lower in
the right plot of figure~\ref{fig-std}, further confirming the
improvement in accuracy when the correct exponent is used.

\begin{figure}

\begin{minipage}{\linewidth}

\includegraphics[width=1\textwidth,height=\textheight]{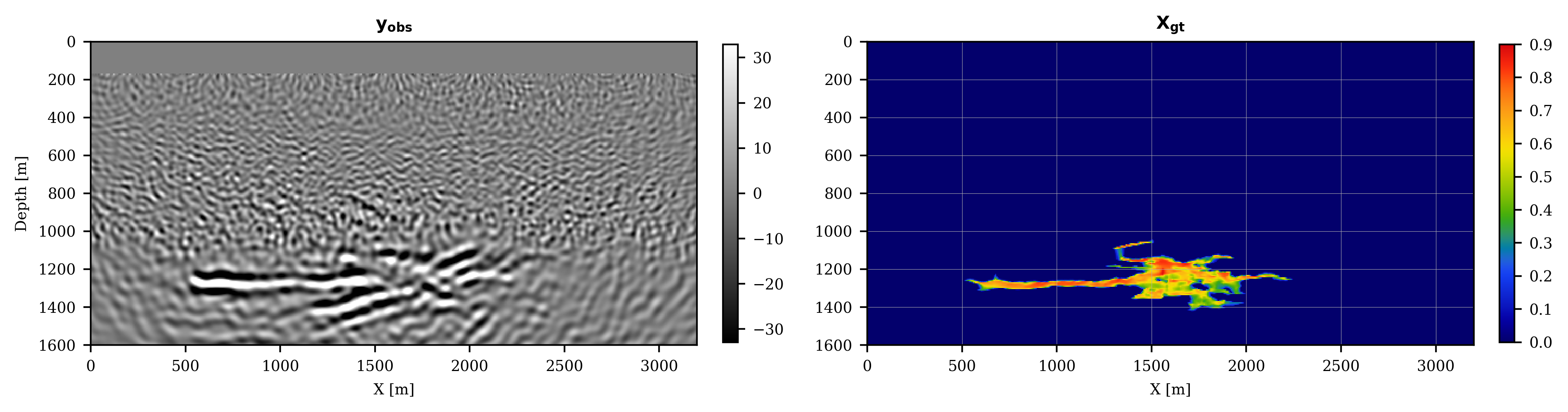}

\end{minipage}%

\caption{\label{fig-GT}Time-lapse seismic observation (left)
corresponding to the ground truth CO\textsubscript{2} plume (right)}

\end{figure}%

\section{CONCLUSIONS}\label{conclusions}

This study demonstrates that uncertainties in subsurface properties,
including permeability and rock physics models, significantly impact the
accuracy of underground energy storage predictions. By incorporating
sensitivity to rock physics variability as a context variable, our
Digital Shadows (DS) accounts for unknown rock physics
models-----particularly within the Brie saturation model family, where
the exponent \(e\) is often unspecified in real-world scenarios. This
context-aware approach enables structured sensitivity analysis, allowing
us to explore ``what if'' questions, for e.g., \emph{what the inference
result would look like if a specific rock physics model were assumed?},
without requiring retraining. By enhancing interpretability and
uncertainty quantification, our method improves the robustness and
reliability of subsurface storage forecasts, supporting more informed
uncertainty-aware decision-making in applications such as hydrogen
storage, compressed air, and geological CO\textsubscript{2} storage.

\section{Acknowledgement}\label{acknowledgement}

This research was carried out with the support of Georgia Research
Alliance, partners of the ML4Seismic Center and in part by the US
National Science Foundation grant OAC 2203821. The overall readability
is enhanced using ChatGPT 4.

\section*{References}\label{references}
\addcontentsline{toc}{section}{References}

\phantomsection\label{refs}

\end{document}